# Ferro-Valleytricity with In-Plane Magnetization

Yibo Liu, Yangyang Feng, Ying Dai[*], Baibiao Huang, and Yandong Ma[*]

School of Physics, State Key Laboratory of Crystal Materials, Shandong University, Shandanan Str. 27, Jinan 250100, China

*Corresponding author: daiy60@sina.com (Y.D.); yandong.ma@sdu.edu.cn (Y.M.)

**Abstract**

Ferro-valleytricity, a fundamental phenomenon that manifests spontaneous valley polarization, is generally considered to occur in two-dimensional (2D) materials with out-of-plane magnetization. Here, we propose a mechanism to realize ferro-valleytricity in 2D materials with in-plane magnetization, wherein the physics correlates to non-collinear magnetism in triangular lattice. Our model analysis provides comprehensive ingredients that allows for in-plane ferro-valleytricity, revealing that mirror symmetry is required for remarkable valley polarization and time-reversal-mirror joint-symmetry should be excluded. Through modulating in-plane magnetization offset, the valley polarization could be reversed. Followed by first-principles, such mechanism is demonstrated in a multiferroic triangular lattice of single-layer $W_3Cl_8$. We further show that the reversal of valley polarization could also be driven by applying electric field that modulates ferroelectricity. Our findings greatly enrich the valley physics research and significantly extend the scope for material classes of ferro-valleytricity.

**Introduction**

Electron valley in condensed-matter physics, characterizing the energy extrema of band, is an internal degree of freedom that resembles spin. Owning to the large separation in momentum space, valley degree is robust against smooth deformation and low-energy phonons, which enables it highly desirable as information carrier [1,2]. To make use of valley degree in information storage and processing, the crucial step is to control the number of electrons in these valleys, thereby generating valley polarization [3-6]. With the recent rise of magnetism in two-dimensional (2D) materials [7-12], the long-sought valley polarization is revealed to be able to occur spontaneously [13]. The discovery of such spontaneous valley polarization in 2D magnetic materials leads to the emergence of the field of ferro-valleytricity, which has attracted broad interest and spurred rapid development in both fundamental research and device applications [13-22].

At present, ferro-valleytricity has been realized in various 2D magnetic systems [13-22]. In all the prior works, it is widely acknowledged that ferro-valleytricity must have an out-of-plane magnetization [14-17]. Physically, using effective model analysis, it is theoretically verified that in-plane magnetization with collinear nature would eliminate the valley polarization induced by spin-orbit coupling (SOC) and thus prohibit ferro-valleytricity [14-17]. On the other hand, as compared with out-of-plane magnetization, in-plane magnetization is more extensive in 2D materials [7,10-12]. Even for the previous identified 2D ferro-valleytronic systems, many of them are only assumed being with out-of-plane magnetization, which actually favor in-plane magnetization [18-21]. Therefore, whether ferro-valleytricity can be realized in 2D materials with in-plane magnetization is an intriguing open issue, which, if realized, will bring a paradigm shift in fundamental physics and candidate system selections of ferro-valleytricity.

In this work, we theoretically predict the realizability of ferro-valleytricity in 2D materials with in-plane magnetization. The physics of the proposed mechanism is rooted in non-collinear magnetism in 2D triangular lattice. Using model and symmetry analysis, the essential ingredients for realizing in-plane ferro-valleytricity are provided, unveiling that mirror (*M*) symmetry is required for remarkable valley polarization and time-reversal-mirror (*TM*) joint-symmetry forbids valley polarization. The resultant valley polarization could be reversed through tuning in-plane magnetization offset. Based on first-principles calculations, this mechanism is also validated in a multiferroic triangular lattice of single-layer $W_3Cl_8$. The reversal of valley polarization is further revealed to be realizable under applying electric field that modulates ferroelectricity. These explored phenomena and insights not only enrich the physics but also expand the material family of ferro-valleytricity.

**Results and Discussion**

To set the stage, we first illustrate the underlying physics that prohibits the formation of ferro-valleytricity with collinear in-plane magnetization. In 2D magnetic lattice with valley physics at K/K' point, SOC plays an dominate role for generating the valley polarization [13-17]. The SOC Hamiltonian of $\mathbf{L} \cdot \mathbf{S}$ are comprised of $H^0$ and $H^1$. $H^0$ describes interaction between the same spin states, and $H^1$ represents interaction between opposite spin states. They can be written as [23,24]:

$$H^0 = \lambda \hat{S}_{z'} \left( \hat{L}_z \cos\theta + \frac{1}{2}\hat{L}_+ e^{-i\phi}\sin\theta + \frac{1}{2}\hat{L}_- e^{+i\phi}\sin\theta \right), \tag{1}$$

$$H^1 = \frac{\lambda}{2}\left( \hat{S}_{+'} + \hat{S}_{-'} \right)\left( -\hat{L}_z \sin\theta + \frac{1}{2}\hat{L}_+ e^{-i\phi}\cos\theta + \frac{1}{2}\hat{L}_- e^{+i\phi}\cos\theta \right). \tag{2}$$

Here, $\hat{L}$ and $\hat{S}$ are orbital and spin angular moments, respectively. The coordinate of $\hat{S}$ is $(x', y', z')$, while the coordinate of $\hat{L}$ is $(x, y, z)$. Polar angles $\theta$ and $\phi$ define the spin orientation, as shown in **Figure S1**. Here, $\hat{L}_\pm = \hat{L}_x + i\hat{L}_y$ and $\hat{S}_{\pm'} = \hat{S}_{x'} + i\hat{S}_{y'}$. As compared with $H^0$, the interaction of $H^1$ can be reasonably neglected [14,17]. In view of the in-plane magnetization with collinear alignment (i.e., $\theta$ is equal to zero), $H^0$ can be simplified as

$$H^0 = \frac{\lambda}{2}\hat{S}_{z'}\left( \hat{L}_+ e^{-i\phi} + \hat{L}_- e^{+i\phi} \right). \tag{3}$$

With considering the $C_{3v}$ group symmetry of K/K' point and all possible basis functions, we obtain the valley polarization as:

$$\Delta E_a = E_a^{K'} - E_a^{K}$$
$$= \langle \varphi^{K'} | H^0 | \varphi^{K'} \rangle - \langle \varphi^{K} | H^0 | \varphi^{K} \rangle = 0. \tag{4}$$

For more detail, please see Note 1 in Supporting Information (SI). Based on Equation (4), it can be easily understood why in-plane magnetization with collinear nature forbids the existence of valley polarization and thus ferro-valleytricity.

Physically, such restriction of in-plane ferro-valleytricity deduced from the above $\mathbf{L} \cdot \mathbf{S}$ theory is based on the paradigm of collinear magnetism. Going beyond the existing paradigm, we present a disruptively new design principle for ferro-valleytricity to tackle this outstanding challenge. The idea is to explore 2D systems in which the valleys correlate to in-plane non-collinear magnetism. **Figure 1a** presents a typical 2D triangular lattice with Y-type non-collinear magnetic order. The magnetic cell consists of three sites where spins lie in the xy-plane forming a 120° structure. To characterize the valley physics of this lattice, we construct an effective tight-binding model for describing the low-energy band dispersions around the Fermi level. The effective Hamiltonian can be expressed as:

$$H = t\sum_{\langle i,j \rangle} c_i^\dagger c_j + m\sum_{\langle i \rangle} c_i^\dagger (\mathbf{s}_i \cdot \boldsymbol{\sigma}) c_i + i\lambda_i \sum_{\langle i,j \rangle} c_i^\dagger (\mathbf{d}_n \times \mathbf{r}_{ij}) \cdot \boldsymbol{\sigma} c_j + i\lambda_r \sum_{\langle i,j \rangle} c_i^\dagger (\mathbf{d}_r \times \mathbf{r}_{ij}) \cdot \boldsymbol{\sigma} c_j. \quad (5)$$

Here, $c_i^\dagger$ ($c_i$) is the creation (annihilation) operator of an electron on site *i*. The first term represents nearest-neighbor hopping with hopping parameter *t*. The second term represents magnetic exchange interaction, wherein *m* is the magnetic moment and $\mathbf{s}_i = (cos(\Delta\varphi), sin(\Delta\varphi), 0)$ corresponds to the in-plane spin vector [as shown in **Figure 1b**]. $\boldsymbol{\sigma}$ is the vector of Pauli matrix. The third and fourth terms describe the intrinsic and Rashba SOC, respectively, with $\lambda_i$ and $\lambda_r$ denoting the strengths. **Figure 1c** shows the origins of SOC, in which the directions of electric fields $\mathbf{d}_n$ (*n* = 1,2,3) that induces intrinsic SOC effect and $\mathbf{d}_r$ that induces Rashba SOC are also indicated.

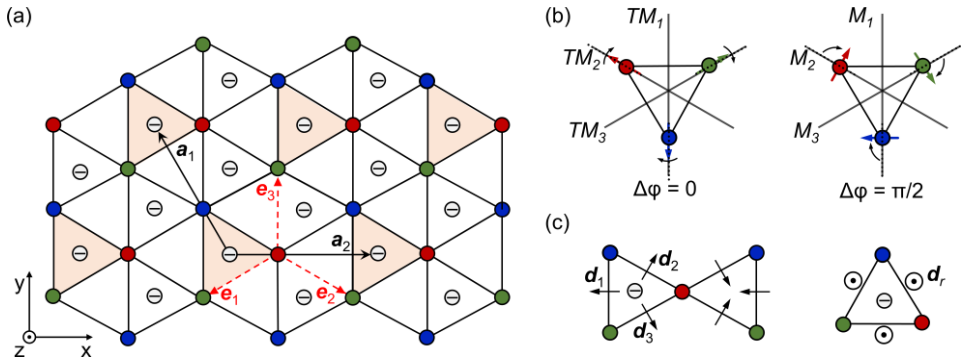

**Figure 1.** (a) Typical 2D triangular lattice with Y-type non-collinear magnetic order. The magnetic cell is highlighted in light orange shadow. Red, blue and green dots represent the three magnetic sites, respectively. $\mathbf{a}_1$ and $\mathbf{a}_2$ are the primitive lattice vectors. Vectors $\mathbf{e}_1$, $\mathbf{e}_2$ and $\mathbf{e}_3$ represent the nearest-neighbor hopping paths from site 1 to site 3. The gray dots located at the center of the triangular lattices represent the ligand ions, which generates electric field that induces SOC. (b) Y-type non-collinear magnetic order with in-plane offset angle $\Delta\varphi = 0$ (left panel) and $\pi/2$ (right panel). The spin

vectors are indicated by arrows. (c) The directions of electric fields $d_n$ ($n$ = 1,2,3) that induces intrinsic SOC effect (left panel) and $d_r$ that induces Rashba SOC (right panel).

The low-energy band dispersions obtained from the tight-binding model for the triangular lattice are depicted in **Figure 2**. When in-plane offset angle $\Delta\varphi$ = 0, $\pi/3$, $2\pi/3$ (**Figures 1b** and **2e**), as shown in **Figure 2b**, the K and K' valleys of the 2D triangular lattice are degenerate in energy, indicating the absence of valley polarization. This is analogy to the case of collinear in-plane magnetization. The underlying physics of such valley degeneracy is sought into the particular magnetic symmetries for the cases of $\Delta\varphi$ = 0, $\pi/3$, $2\pi/3$, i.e., the *TM₁, TM₂,* and *TM₃* joint-symmetries; see **Figures 1b** and **2e**. Here, we use $\mathbb{O}$ instead of *TM*, and select two Bloch states denoted as $|\varphi(\boldsymbol{k})\rangle$ and $|\phi(\mathbb{O}\boldsymbol{k})\rangle=\mathbb{O}|\varphi(\boldsymbol{k})\rangle$ with energies E($\boldsymbol{k}$) and E($\mathbb{O}\boldsymbol{k}$), respectively. According to the symmetry requirement

$$\mathbb{O}H(\boldsymbol{k})\mathbb{O}^{-1}=H(\mathbb{O}\boldsymbol{k}), \tag{7}$$

we have

$$H(\mathbb{O}\boldsymbol{k})|\phi(\mathbb{O}\boldsymbol{k})\rangle =\mathbb{O}(H(\boldsymbol{k})|\varphi(\boldsymbol{k})\rangle)=E(\mathrm{k})|\phi(\mathbb{O}\boldsymbol{k})\rangle. \tag{8}$$

Namely, the energies in momentum space along the Γ-K line satisfy the relation E($\boldsymbol{k}$) =E(-$\boldsymbol{k}$), subsequently suppressing polarization at the K and K' valleys [25].

Evidently, to induce possible valley polarization in such 2D triangular lattice with Y-type non-collinear magnetic order, the *TM₁, TM₂,* and *TM₃* joint-symmetries must be excluded. In principle, this can be readily achieved through modulating the in-plane offset angle $\Delta\varphi$. Indeed, our model analysis shows that with breaking the three *TM* joint-symmetries through tuning in-plane offset angle $\Delta\varphi$, the K and K' valleys are no longer degenerate in energy, thereby leading to the appearance of spontaneous valley polarization and thus the long-sought in-plane ferro-valleytricity. For example, as shown in **Figure 2b**, for in-plane offset angles $\Delta\varphi$ = $\pi/6$, $\pi/2$ (**Figures 1b** and **2e**), significant valley polarizations are observed. More intriguingly, these two cases both exhibit the unique single valley state that is observed in out-of-plane magnetization system [26].

Based on the tight-binding model analysis, we also find that the value of the realized valley polarization significantly depends on the SOC parameter $\lambda_r$. Taking in-plane offset angle $\Delta\varphi$ = $\pi/2$ as an example, we can see that when $\lambda_r$ = 0, the valley polarization vanishes (**Figure 2c**). With increasing $\lambda_r$ from 0 to 0.3$t$, the valley polarization monotonically increases and becomes significant. As significant valley polarization is favorable for ferro-valleytricity, SOC parameter $\lambda_r$ should be increased as large as possible. To increase $\lambda_r$, one can enlarge the electric field $d_r$, i.e., significantly breaking the out-of-plane $M$ symmetry, and introduce heavy metal atoms to enhance the SOC strength.

To obtain a more comprehensive understanding of the effect of $\lambda_r$ and $\Delta\varphi$ on the valley polarization, we plot the 3D surface plot and 2D counter map of valley polarization as functions of these two parameters in **Figure 2d**. It can be clearly seen that except for the in-plane offset angles $\Delta\varphi = 0, \pi/3, 2\pi/3$, the valley polarization emerges, which relates to the broken of $TM_1$, $TM_2$, and $TM_3$ joint-symmetries. Intriguingly, for a given $\lambda_r$, the valley polarization reaches a peak under the in-plane offset angles $\Delta\varphi = \pi/6, \pi/2$. As shown in **Figures 1b** and **2e**, these two cases correspond to the existence of in-plane $M$ symmetry. As a result, for the in-plane ferro-valleytricity realized in such 2D triangular lattice, the $M$ symmetry is required for significant valley polarization. Another important feature we can see from **Figures 2b** and **2d** is that the valley polarization under the in-plane offset angle $\Delta\varphi$ is opposite to that under $\Delta\varphi \pm \pi/3$. This indicates that the valley polarization of the in-plane ferro-valleytricity could be reversed by tuning in-plane magnetization offset. As comparison, it is interesting to point out that such valley polarization reversal in ferro-valleytricity with collinear magnetization is realized through inversing spin orientation [13].

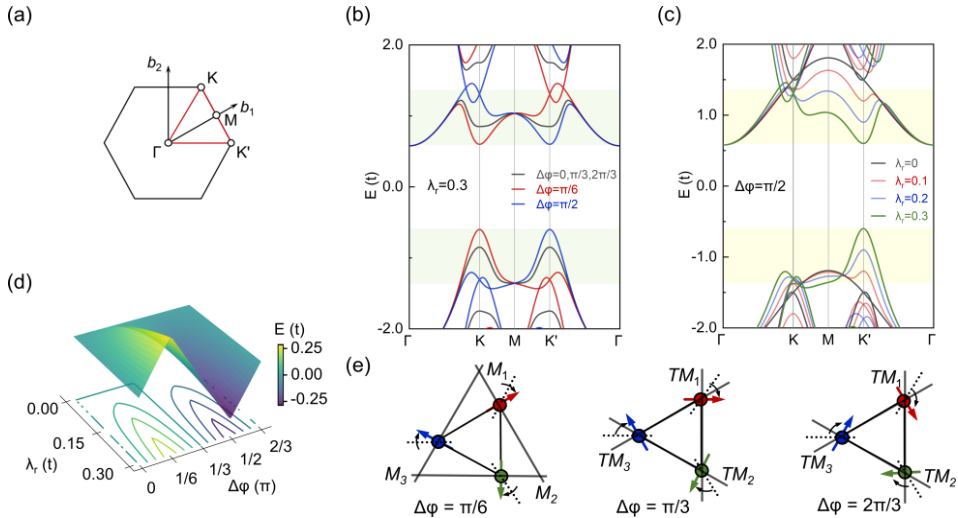

**Figure 2.** (a) 2D Brillouin zone with high-symmetry points. Band dispersions obtained from the tight-binding model for the triangular lattice with (b) different in-plane offset angles, $m = 1.5t$ and $\lambda_i = 0.4t$ and (c) different $\lambda_r$, $\Delta\varphi = \pi/2$, $m = 1.5t$ and $\lambda_i = 0.4t$. (d) 3D surface plot and 2D counter map of valley polarization as functions of $\lambda_r$ and $\Delta\varphi$, with $m = 1.5t$, $\lambda_i = 0.4t$. (e) The corresponding coplanar

magnetic textures in (b).

One candidate material for realizing this physics is single-layer (SL) $W_3Cl_8$. The space group of SL $W_3Cl_8$ is $P3m1$ (No. 156). Its lattice parameter is optimized to be 6.66 Å, which is consistent with previous work [27]. As depicted in **Figure 3a**, in the vertical direction, there is a W atomic layer in the middle, with two asymmetric Cl atomic layers on the outside. Each W atom is surrounded by six Cl atoms, forming a twisted octahedral structure with shared edges. The W atoms are arranged in a kagome-like lattice, and every three adjacent W atoms form a $W_3$ trimer, resulting in a breathing kagome lattice [27,28]. Notably, as depicted in **Figure 3b**, the $W_3$ trimers form a triangular lattice, and the Cl atoms exhibit an asymmetric distribution. Consequently, both the inversion symmetry and out-of-plane $M$ symmetry of SL $W_3Cl_8$ are broken, leading to electrical polarization along the out-of-plane direction.

The valence electronic configuration of W is $5d^4 6s^2$. Prior to forming a trimer, the d orbitals of W split into $e_g$ and $t_{2g}$. The formation of a trimer further causes the d orbitals to split into $1e$, $1a_1$, $2e$, $2a_1$, $3e$ and $1a_2$. Here, $e$, $a_1$ and $a_2$ are irreducible representations of the $C_{3v}$ point group, where $e$ is the basis function of $(d_{x^2-y^2}, d_{xy})$, $(d_{xz}, d_{yz})$, and $a_1$ is the basis function of $(d_{z^2})$. The three W atoms share ten electrons in one trimer, which results in a magnetic moment of $2\mu_B$, as illustrated in **Figure 3c**. Consequently, the trimer is spin-polarized. Moreover, our calculations show that it prefers in-plane magnetization with a magnetocrystalline anisotropy energy of -28 meV per unit cell. To estimate the magnetic ground state of SL $W_3Cl_8$, we consider nonmagnetic, ferromagnetic and three types of antiferromagnetic configurations. The Y-type antiferromagnetic configuration is found to have the lowest energy, which satisfies the ingredients required by the proposed mechanism. The details of these results are presented in the SI.

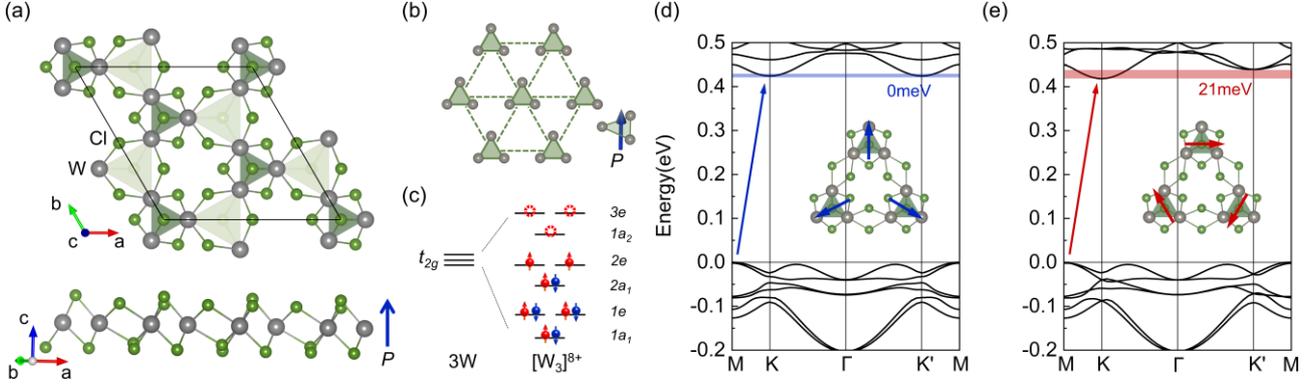

**Figure 3.** (a) Crystal structure of SL $W_3Cl_8$. (b) Triangular lattice formed by $W_3$ trimers and the resultant electric polarization. (c) Splitting of *d*-orbitals of $W_3$ trimers, with red and blue arrows representing spin-up and spin-down electrons, respectively. (d) Band structure of SL $W_3Cl_8$ with SOC under in-plane offset angle $\Delta\varphi = 0$. (e) Band structure of SL $W_3Cl_8$ with SOC under in-plane offset angle $\Delta\varphi = \pi/2$. Insets in (d) and (e) depict the corresponding magnetic textures. The Fermi level is set to VBM.

The band structures of SL $W_3Cl_8$ with SOC under in-plane offset angles $\Delta\varphi = 0$ and $\pi/2$ are presented in **Figure 3d** and **3e**, respectively. It is evident that SL $W_3Cl_8$ displays an indirect bandgap semiconductor character, with its valleys in the conduction band minimum locate at the K and K' points. As depicted in **Figure 3d**, when in-plane offset angle $\Delta\varphi = 0$, the valleys are degenerate in energy, indicating the absence of valley polarization. When the in-plane offset angle $\Delta\varphi$ is adjusted to $\pi/2$, as illustrated in **Figure 3e**, the valley degeneracy is broken, leading to a valley polarization of 21 meV. This value is larger than that observed in out-of-plane ferro-valleytronic systems of $Cr_2Se_3$ and $VAgP_2Se_6$, and comparable to that of $LaBr_2$ [29-31]. Consequently, the realizabilities of ferro-valleytricity in SL $W_3Cl_8$ with in-plane magnetization is demonstrated, agreeing well with our proposed mechanism.

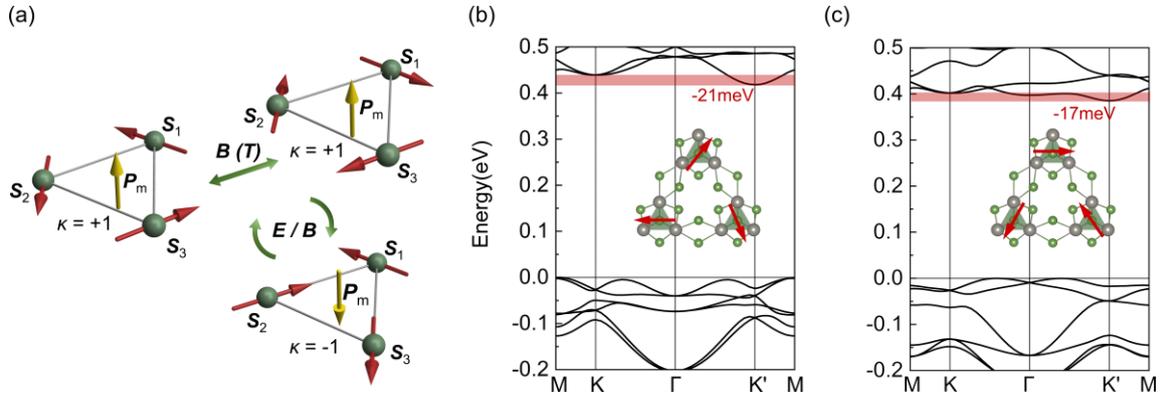

**Figure 4.** (a) Diagram illustration of the switching between different valley states by modifying the magnetic texture through the application of magnetic or electric fields. Yellow arrows centered within the triangular lattices represent the ferroelectric polarizations induced by noncollinear magnetic texture. (b) Band structure of SL $W_3Cl_8$ with SOC in-plane offset angle $\Delta\varphi = \pi/6$. (c) Band structure of SL $W_3Cl_8$ with SOC under in-plane offset angle $\Delta\varphi = \pi/2$ and opposite vector-spin chirality. The Fermi level is set to the VBM.

Based on the proposed model, the in-plane valley polarization can be reversed by altering the in-plane offset angle. As illustrated in the top panel of **Figure 4a**, the in-plane offset angle can be altered through the application of a magnetic field. With adjusting the in-plane offset angle $\Delta\varphi$ from $\pi/2$ to $\pi/6$, the valley polarization is reversed from 21 to -21 meV; see **Figures 3e** and **4b**, indicating the successful realization of magnetic manipulation of in-plane ferro-valleytricity. Furthermore, due to existence of the electric polarization driven by the chirality of the magnetic vectors, the chirality of the in-plane magnetization can be reversed by applying electric field [32,33], as illustrated in the bottom panel of **Figure 4a**. With reversing the chirality of SL $W_3Cl_8$ under in-plane offset angle $\Delta\varphi = \pi/2$, as shown in **Figures 3e** and **4c**, the valley polarization is adjusted from 21 to -17 meV. Therefore, the reversal of valley polarization is further revealed to be realizable under applying electric field that modulates ferroelectricity.

**Conclusion**

In conclusion, a mechanism of realizing ferro-valleytricity in 2D materials with in-plane magnetization is proposed. The physics is related to non-collinear magnetism in triangular lattice. The comprehensive ingredients that allows for in-plane ferro-valleytricity is provided, indicating that *M* symmetry is required for remarkable valley polarization and *TM* joint-symmetry should be excluded. By adjusting in-plane magnetization offset, the valley polarization could be reversed. Using first-principles, this mechanism is further demonstrated in a multiferroic triangular lattice of single-layer $W_3Cl_8$. We also find that the reversal of valley polarization could also be driven by applying electric field that modulates ferroelectricity.

**Methods**

Our first-principles calculations are carried out based on density functional theory (DFT), implanted in Vienna *ab initio* Simulation Package (VASP) [34,35]. The ionic potential is described by the projected augmented wave (PAW) approach [36]. The exchange-correlation interactions are characterized by the Perdew-Burke-Ernzerhof (PBE) functional of generalized gradient approximation (GGA) [37]. To avoid interactions between adjacent layers, a vacuum layer of more than 15 Å is set along the out-of-plane direction. A 5×5×1 Monkhorst-Pack k-point mesh is adopted to sample the 2D Brillouin zone [38]. The cutoff energy and the convergence criterion are set to 500 and $10^{-6}$ eV, respectively. All atoms are fully relaxed until the atomic forces are less than 0.01 eVÅ$^{-1}$. SOC is considered in the electronic structure calculations.

**Supporting Information**

The Supporting Information is available free of charge at ***.
(Figure S1) Polar angles $\theta$ and $\phi$ defining the relative orientation between spin and orbital coordinate systems; (Note 1) The detailed derivation of energy differences for different p and d orbitals as basis functions. (Figure S2) Schematic of magnetic states for SL $W_3Cl_8$.; (Table S1) The energies of the SL $W_3Cl_8$ for different magnetic states.

**Notes**

After we finishing the preparation of this work, we find a similar and independent work is published [39]. However, the tight-binding model and candidate material we used are significantly different from Ref. [39].

**Acknowledgements**

This work is supported by the National Natural Science Foundation of China (Nos. 12274261 and 12074217), Shandong Provincial QingChuang Technology Support Plan (No. 2021KJ002), and